# Gaussian pulse-induced imperfect rephasing in coherent transients

**Byoung S. Ham**[*]

Center for Photon Information Processing, and School of Electrical Engineering and Computer Science, Gwangju Institute of Science and Technology
123 Chumdangwagi-ro, Buk-gu, Gwangju 61005, S. Korea

**Abstract.** A double-rephasing photon-echo (DRPE) scheme inherently satisfies no-population inversion condition for quantum memory applications, while the resultant absorptive coherence prohibits echo radiation out of the medium. To solve the absorptive echo problem in the DRPE scheme, a controlled double-rephasing echo protocol has been presented and discussed for the collective atom phase control. In the DRPE scheme, however, a typical Gaussian light can cause imperfect rephasing processes resulting in a partial emission of photon echoes. Here, the rephasing process of individual ensemble atoms is thoroughly investigated to discuss the unpermitted echo observations in the DRPE scheme, where its maximum efficiency is as high as 26% in amplitude.
(Submitted on May 07, 2017; Version 3)

## I. INTRODUCTION

Quantum memory [1-9] has become an emerging research topic of quantum information area recently faced with serious challenges regarding qubit scalability [10] due mainly to the short decoherence time of qubits. The qubit scalability plays an important role not only for quantum algorithms such as Shor's prime number factorization [11] and Grover's data search [12], but also for fault-tolerant quantum computers [13] and long-haul quantum communications via quantum repeaters [14]. For the fault-tolerant quantum computing not only fidelity but also retrieval efficiency plays an important role to satisfy a minimum level of quantum error corrections [15]. For the quantum communications imperfect devices may allow potential eavesdropping, and so does the low retrieval efficiency of quantum memories. Thus, both storage time and retrieval efficiency of quantum memories become major parameters to determine the functionality of quantum technologies.

Over the last decade, photon echoes [16] have been intensively studied as a potential candidate of quantum memories owing to the inherent benefits of ultrafast, wide bandwidth, and multi-mode information processing capabilities [1-7]. The key mechanism of photon echoes is in the classical controllability of ensemble coherence. The classical controllability is achieved by an optical π pulse, resulting in the reversible coherence evolution of the ensemble. However, the optical π pulse induces a complete population inversion for especially weak data pulses, causing inevitable quantum noises to the retrieved signals due to spontaneous and/or stimulated emissions. Unlike classical information processing, quantum information does not allow duplication (or cloning) of an unknown quantum state [17]. Thus, the population inversion in photon echoes has been a fundamental constraint to be overcome for quantum memory applications [1-7].

A double-rephasing photon-echo (DRPE) scheme has been adapted for quantum memories due to its inherent property of no population inversion [18-23]. Although the first echo in the DRPE scheme has been well treated to be silent, so that it does not affect the second (final) echo [18-20], the inherent absorptive coherence of the second echo prohibits echo generations out of the medium [21]. To solve this absorptive echo problem in the DRPE scheme, a controlled double rephasing (CDR) echo protocol has been proposed, where controlled coherence conversion (CCC) plays a key role for the ensemble phase control [21-23]. The CCC uses a controlled Rabi flopping via a population oscillation between the excited state in the DRPE scheme and a third state. Unlike a two-level system of photon echoes including DRPE, however, the three-level system of CDR gains a π phase shift for a complete Rabi flopping by CCC [24]. Such an optical phase gain has been firstly observed and discussed in resonant Raman echoes [25] and applied to photon echo-based quantum memories [1]. Thus, CCC has become a key mechanism to manipulate the absorptive coherence of DRPE. Recently observed DRPE without CCC [18-20], however, seems to contradict the CDR echo theory. Here, I thoroughly investigate rephasing process by a typical Gaussian light pulse and identify that the observed DRPEs [18-20] are not contradiction to the CDR echo theory but defects resulting from imperfect rephasing by the Gaussian light.

## II.     ANALYSIS

Due to the Gaussian light profile $G_j$, a commercial laser light pulse can simultaneously generate nearly all kinds of pulse areas ($\sum \Phi_j$), where $\Phi_j = \int_0^T \Omega_j \, dt \cong \Omega_j T$. Here, the Rabi frequency $\Omega_j$ depends only on $G_j$, where $j$ is the position along the transverse direction with respect to the pulse propagation direction, and T is the time duration of a square pulse which can be easily obtained by modern electro-optic devices. To analyze the echo observations in DRPE schemes [18-20] seemingly contradicting the CDR echo theory [21-23], ensemble coherence evolutions are numerically investigated for individual atoms interacting with the Gaussian distributed light. The axial distribution of the light pulse is excluded, because square pulses can be easily manipulated. To fully visualize the atom coherence evolutions, all decay rates of the medium are set to be zero. Thus, the phase evolution of the ensemble coherence relies on both optical inhomogeneous broadening of the ensemble and nonuniform spatial distribution of the Gaussian light.

For the present analysis, time-dependent density matrix equations are numerically solved for a two-level ensemble medium interacting with resonant optical pulses in the Heisenberg picture under rotating wave approximations [26]: $\frac{d\rho}{dt} = \frac{i}{\hbar}[H,\rho] - \frac{1}{2}\{\Gamma,\rho\}$, where ρ is a density matrix element, H is interaction Hamiltonian, and Γ is a decay parameter. For this a commercial laser light whose spatial profile is Gaussian distributed is taken as a light source. The Gaussian light induces a Gaussian profile of interacting atoms, where the atom distribution is divided into 41 subgroups to cover 99.55% of the total distribution. Each subdivision of the Gaussian light-interacting atoms is divided into 281 spectral groups at 10 kHz spacing for 1.2 MHz (FWHM) inhomogeneous broadening. Those 281 spectral groups are individually calculated for the time-dependent density matrix equations and summed for all spatial components of the Gaussian distributed atom groups, unless otherwise specified. The Gaussian light-corresponding atom groups do not overlap (or interact) with each other, resulting in no interference among them due to transverse (spatial) distribution. So do the photon-echo components due to the phase matching condition: $\boldsymbol{k}_{echo} = 2\boldsymbol{k}_{rephasing} - \boldsymbol{k}_{data}$. Thus, the sum for all 41 Gaussian-distributed atom groups denotes the overall ensemble coherence evolutions. In the DRPE scheme, only the emissive components in the second photon echo contribute to the echo observations.

The optical pulse duration is set to be 0.1 μs, and the time increment in the calculations is also 0.1 μs. Initially all atoms are in the ground state $|1\rangle$ ($\rho_{11} = 1$), and thus all initial coherences are $\rho_{ij} = 0$, where $i \neq j$. The present numerical simulations are time-interval independent, so that there is no accumulated error dependent upon the time interval settings. For a π pulse area the corresponding Rabi frequency of the 0.1 μs–pulse is 5 MHz multiplied by 2π. From now on the 2π multiplication factor is omitted for simplicity. Here, the pulse area of a Gaussian light pulse in the transverse mode can be described with the Gaussian distribution $G_j$. Unlike Maxwell-Bloch approach, the present numerical calculations can show the details of coherence evolutions in time domain without approximations. This is the essential benefit of the present numerical method for photon echo analyses.

Figure 1 shows conventional two-pulse photon echo simulation results, where the light pulse(s) has a Gaussian profile in a transverse direction as mentioned above. The spatial magnitude of the light intensity perfectly maps onto the cross section of an interacting ensemble, resulting in a Gaussian distribution of atom groups. For simplicity of symmetry, we take only one-dimensional transverse mode of the cross section. For the analysis of the present Gaussian pulse-based photon echoes, three types of two-pulse combinations can be considered: The first column in Fig. 1 is for a Gaussian pulse applied to the data (D) only; The second column is for a Gaussian pulse applied to the rephasing (R) only; The third column is for Gaussian pulses applied to both D and R. The first row of Fig. 1 shows the three different cases of pulse combinations. The second row shows the corresponding density matrix calculations of the first row, where the complete rephasing appears at t=4.1 μs (dotted line). The third row shows the peak amplitude (red open circles) of the photon echoes in the second row along the dotted line, where the solid line represents for the theory of the best-fit curve.

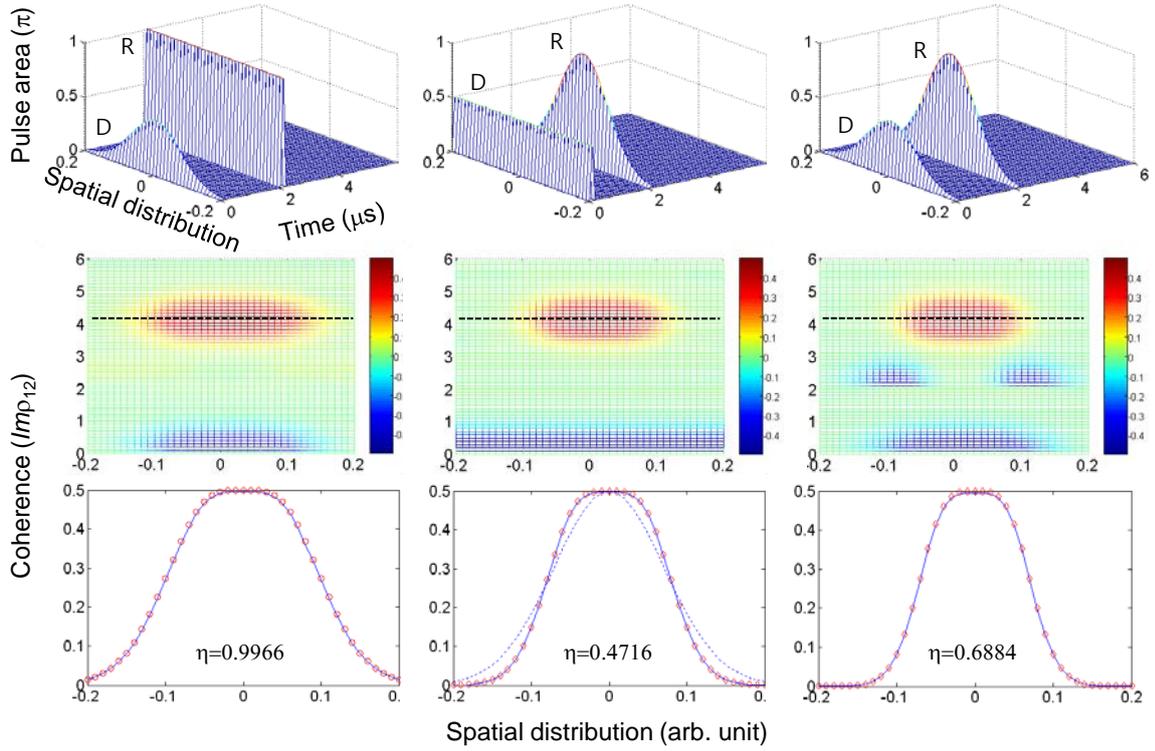

**Figure 1.** Gaussian pulse-dependent two-pulse photon echoes. Top row: three-different combinations of Gaussian pulse applications; Middle row: corresponding results of the top row; Bottom row: corresponding photon echo efficiency η. The arrival times of data (D) and rephasing (R) pulses are $t_D$=0.1 μs and $t_R$=2.1 μs, respectively. The Gaussian distribution is along the transverse direction of the beam propagation. The blue curve in the bottom raw indicates a theoretical fit, where the red circles are for the corresponding echo signals at t=4.1 μs along the dotted line in the Middle row. From left to right in the Bottom row, the theoretical fit curves are $sin(G_j)$, $sin^2(G_j)$, and $sin^3(G_j)$, where $G_j$ is the Gaussian distribution in the Top row. Dotted curve: the Gaussian profile. The echo efficiency η is defined by $\frac{|\Sigma_j \rho_{12}(t_e)|}{|\Sigma_j \rho_{12}(t_d)|}$, where j is the 41 groups in the Gaussian distribution, and $t_d$ ($t_e$) is the data (echo) arrival time for maximum coherence.

For the first case of the Gaussian D in the first column, the echo in the second and third rows perfectly fits the function of *sin(G_j)*, where $G_j$ represents the Gaussian profile. This exactly mimics the area theorem of photon echoes [27,28], where an optical π pulse perfectly rephases all of the data components. Although the area theorem originates in the time domain, the echo efficiency η reaches 100% as shown in the lower left corner (see the Supplementary information Figure S1). Here, the D-pulse-induced coherence profile perfectly overlaps with the echo profile.

For the second case of the Gaussian R in the second column, the theoretical best-fit curve is *sin²(G_j)*. Because the π rephasing pulse is considered as a double coherence excitation of D, the square law is quite reasonable. Here all components of echo efficiency are lower than unity except for the π/2−π pulse sequence at line center of $G_j$. The resulting echo efficiency is cut by ~50%. In fact, the echo efficiency can be increased by expanding the width of the Gaussian profile up to 100%, in which 100% echo efficiency corresponds to a uniform rephasing pulse as shown in the first column.

For the last case of both Gaussian pulses in the third column, the echo distribution profile is shrunken narrower than that the coherence excited by the Gaussian D pulse, where the echo efficiency drops down to ~70%. Interestingly the best-fit curve is *sin³(G_j)*, which is the product of the first and second cases. Here, the Gaussian pulse can be interpreted to induce a 30% coherence leakage due to imperfect rephasing process by the Gaussian rephasing pulse, and this leakage allows unpermitted echo generations in the DRPE of refs. 18-20 (will be discussed in Figs. 2 and 3). From the analyses in Fig. 1, the Gaussian rephasing pulse results in imperfect rephasing process-caused coherence leakage.

## III. RESULTS AND DISCUSSIONS

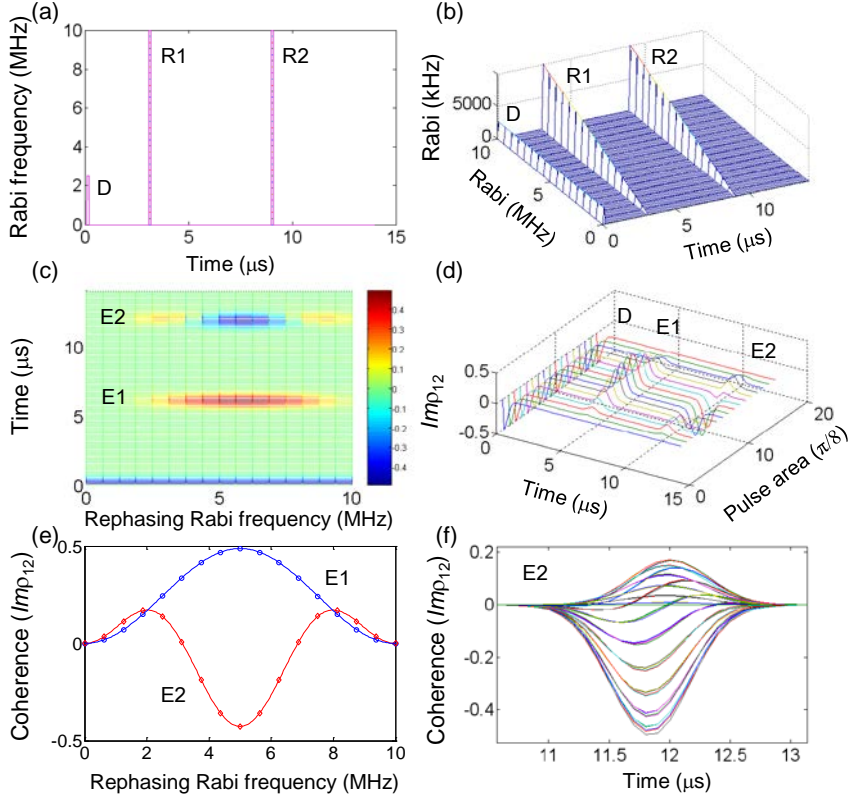

**Figure 2.** Doubly rephased photon echoes for linearly varying rephasing pulse area. (a) Pulse sequence for the double rephasing photon echo. D: Data pulse ($\pi/2$, fixed); R1: First rephasing pulse; R2: Second rephasing pulse. (b) Spatially varying rephasing Rabi frequency for (a). The 10 MHz Rabi frequency corresponds to $2\pi$ pulse area. (c) Corresponding results for (b). E1 (E2) represents the first (second) echo. (d) Individual atom phase evolutions for each varying Rabi frequency of the rephasing pulses in (c). (e) Rephasing Rabi frequency-dependent photon echo amplitudes of E1 (blue circles) at t=6.1 μs and E2 (red diamonds) at t=12.1 μs: see equations (1) and (2) in the text. Each colored curve is the best-fit theory (see the text). (f) Overlapped E2 in (d).

Now let's consider the unpermitted DRPEs observed recently, where the first echo is erased macroscopically [18-20]. Figure 2(a) shows a typical pulse sequence of the DRPE scheme composed of identical double rephasing pulses. Here, the first echo generation is assumed to be silent, where this silencing process does not affect individual coherence evolutions [18]. As discussed in refs. 21-24, the second echo generation out of the medium in DRPE is strongly prohibited due to its absorptive coherence, if spatially uniform optical $\pi$ pulses are used. Then, the DRPE observations must be due to emissive components resulted from the imperfect rephasing process in Fig. 1. Figure 2(b) shows linearly varying rephasing pulse profile of R1 and R2, where this variation is spatially distributed in a transverse mode. This spatial variation of the rephasing pulses R1 and R2 corresponds to the pulse area varying from zero to $2\pi$ (10 MHz in Rabi frequency) for a fixed $\pi/2$ pulse area of D.

Figures 2(c) and 2(d) represent corresponding results of Fig. 2(b), where emissive components of E2 are generated when the rephasing pulse area falls below $\sim\pi/2$ or above $\sim 3\pi/2$. This is obviously due to the imperfect rephasing by R1 as discussed in Fig. 1. Figure 2(e) is the plot of echo amplitudes in Fig. 2(c) at each echo timing as a function of the rephasing pulse area. The echo E2 (red open circles) has certain regions of emissive components $(Im\rho_{12} > 0)$. These emissive components of E2 are due to imperfect rephasing in E1, where the rephasing pulse area $\Phi_R$ satisfies one of the following inequalities:

$$2n < \frac{\Phi_R}{\pi} < 2n + \alpha, \quad (1)$$

$$2(n+1) - \alpha < \frac{\Phi_R}{\pi} < 2(n+1), \quad (2)$$

where $\alpha$ is 5/8, and n=0,1,2,… The $\Phi_R$ equally applies to both R1 and R2 (see also the Supplementary Figure S2). Each solid curve represents the best-fit curve for the data in Fig. 2(c), where the best-fit curves (echo amplitudes) for E1 and E2 are intuitively obtained as:

$$E1 = sin^2\left(\frac{\Phi_R}{2}\right)/2, \quad (3)$$

$$E2 = -\sqrt{2}\left[sin^2\left(\frac{\Phi_R}{2}\right)[0.3 - cos^2\left(\frac{\Phi_R}{2}\right)]\right]. \quad (4)$$

The first echo E1 is similar to the second case (second column) in Fig. 1. Here the condition for the silent echo E1 is satisfied without violating the generality of the present analysis, because the silence does not alter individual coherence evolutions [18]. Thus, we conclude that the final echo E2 in the DRPE scheme can be observed if the rephasing pulse area $\Phi_R$ satisfies equation (1) or (2). This condition is theorized by equation (4) with E2>0. As a result, the echo observations in the DRPE experiments of refs. 18-20 are due to imperfect rephasing by Gaussian light pulses without contradiction to the CDR echo theory [22]. Figure 2(f) shows all components of the final echo E2 overlapped in time domain.

Figure 3 shows a real case of the DRPE, where all the light pulses interacting with a medium are Gaussian distributed along the transverse direction with respect to the beam propagation direction. The Gaussian spatial distribution of the light in Fig. 3(a) is divided into 41 groups, where each group interacts with independent atoms whose inhomogeneouse broadening is 1.2 MHz (FWHM). The data pulse is assumed weak: $\Phi_D = \pi/5$. The two red open circles in Fig. 3(a) indicate the critical value calculated by equation (4) for E2=0. Figures 3(c) and 3(d) show the results of individual coherence evolutions for Fig. 3(b). In Fig. 3(d), the second echo E2 shows partially emissive components, satisfying equation (1) (see the dotted circles). In Fig. 3(e), the first echo E1 at t=6.1 μs, and the second echo E2 at t=12.1 μs are shown with respect to the Gaussian distributed spatial mode. The red curve represents only for the emissive components E2$_{eff}$ ($Im\rho_{12}$>0 in E2) selected from the second echo E2 represented by the dashed curve. The emissive regions in E2$_{eff}$ also satisfy equation (1) for $\alpha \cong \frac{5\pi}{8}$. Here the negative components of E2 does not alter the final echo generation E2 due to complete independence of the interacting atoms determined by the spatial distribution of the Gaussian pulse in Fig. 3(a). Figure 3(f) shows sum coherence evolution for E2$_{eff}$ in Fig. 3(e) in time domain. The echo efficiency η (E2$_{eff}$/D) in Fig. 3(f) is 6.9 %. For η, peak amplitude of each coherence is compared, where the actual data-induced coherence is bigger than the appeared because Fig. 3(f) is only for $Im\rho_{12} > 0$ at the second echo timing in Fig. 3(d). It should be emphasized that the first echo E1 has been treated not to be macroscopically rephased [18] even if it is appeared in the present calculations.

Figure 3(g) is accumulations of Fig. 3(f) for different peak pulse areas of the Gaussian rephasing pulse in Fig. 3(b), where the peak pulse area of the rephasing Gaussian profile varies from zero to 2π at a π/4 step, by increasing the rephasing Rabi frequency $\Omega_R$ (=$\Omega_{R1}$=$\Omega_{R2}$) (see also the Supplementary Figure S3). As mentioned in Fig. 3(f), the data pulse-induced coherence is appeared in different magnitudes only because of the different amounts of emissive components for E2 for $Im\rho_{12} > 0$. In Fig. 3(h), echo efficiency η $\left(\left|\frac{E2_{eff}}{D}\right|\right)$ is plotted for different rephasing pulse area in Fig. 3(g) (see the dotted circle), where η shows a damped oscillation as the rephaisng pulse area increases. The damping is due to inhomogenously broadened atoms as discussed in ref. 23. From Fig. 3(h), it is concluded that the unpermitted photon echo E2 in the DRPE scheme always exists regardless of the rephasing pulse area if the rephasing pulses are Gaussian distributed. The maximum echo efficiency of 26% is achieved when the rephasing pulse area is π/2 at Gaussian peak. The η converges at ~10% as the rephasing pulse area increases (see the best-fit curve). Figure 3 now clearly explains how unpermitted echo observations in the DRPE schemes (in refs. 18-20) were possible, showing that it is a defect by Gaussian rephasing pulses. In other words, the unpermitted echo generation in the DRPE scheme of Fig. 3(b) is understood as a coherence leakage due to the imperfect rephasing by Gaussian pulses. The low efficiency of DRPE as shown in Fig. 3(h), however, has no direct relation with fidelity as long as spontaneous or stimulated emission is not involved.

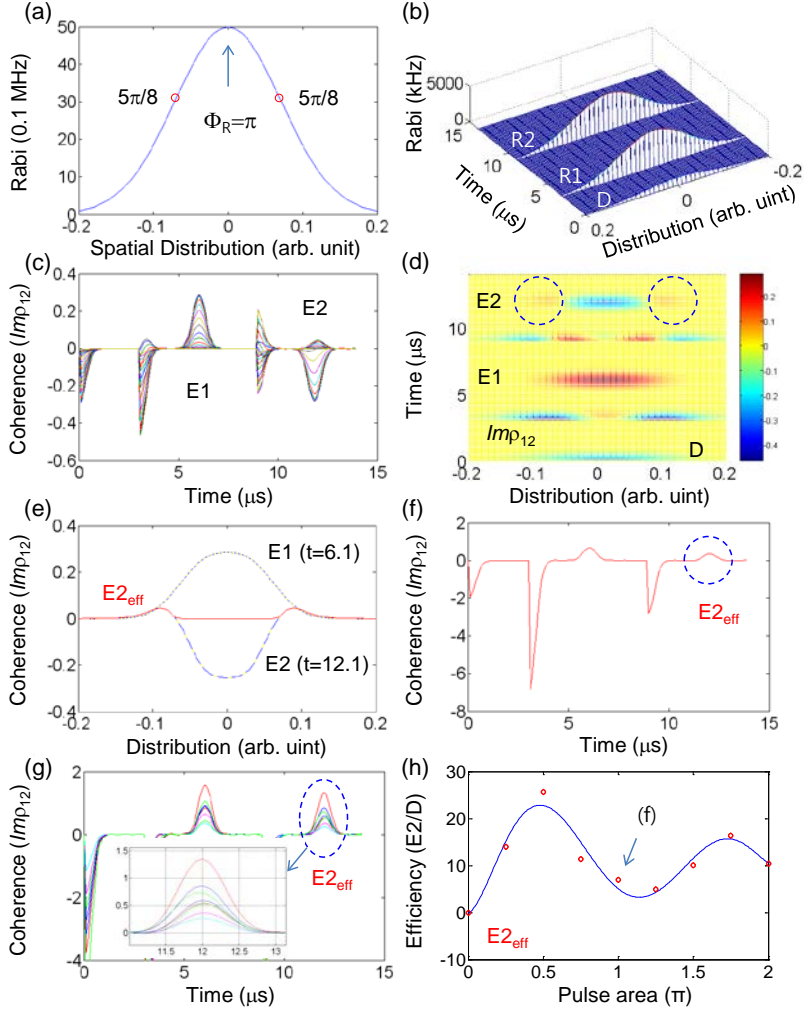

**Figure 3.** Gaussian pulse induced photon echoes in a double raphaisng scheme. (a) Gaussian spatial profile of the rephasing pulse R1 and R2. The peak power (Rabi frequency) at center corresponds to a π pulse area. Red circles indicate the pulse area satisfying equations (1) and (2) in the text. The Gaussian spatial profile of (a) is divided into 41 groups at 10 kHz spacing, as shown in (b). (b) Pulse sequence for a double rephasing photon echo. Data pulse area $\Phi_D$ is fixed at $\Phi_D=\pi/5$. The arrival time of Data D, Rephsing R1, and R2 is 0.1, 3.1, and 9.1 μs, respectively. (c) Individual coherence evolutions of $Im\rho_{12}$ for all 41 components in (b). (d) A 3D color map for (c). (e) Coherence ($Im\rho_{12}$) as a function of Gaussian distribution for E1 and E2 in (d). The red curve is for E2$_{eff}$, where $Im\rho_{12} > 0$. (f) Sum of E2$_{eff}$ for all Gaussian distribution in (e). (g) Overlapped sum coherene $Im\rho_{12}$ for different intensities of the Gaussian rephasing pulses in (b): for the peak pulse area ($\Phi_R$) in (a), Green: $\pi/4$; Red: $\pi/2$; Blue: $3\pi/4$; Magenta: $\pi$; Cyan: $5\pi/4$; Green: $3\pi/2$; Blue: $7\pi/4$; Red: $2\pi$. The reason of different magnitude in D-induced coherence is due to E2$_{eff}$ (see the text). (h) Red circle: Echo efficiency η $\left(\left|\frac{E2_{eff}}{D}\right|\right)$ in (g), where D stands for $\rho_{12}(t_d)$. Blue line: best-fit curve for the data (red circles). The arrow mark is for (f). The unit of coherence in (f) and (g) is arbitrary. The unit of efficiency in (h) is %.

## IV. CONCLUSION

In conclusion, rephasing processes in a double-rephasing photon-echo (DRPE) scheme were thoroughly investigated for the ensemble coherence evolutions and the DRPE observations were identified as a result of imperfect rephasing processes by the Gaussian light pulses. The theoretical formula for the unpermitted echo observations in DRPE was also induced. As a result there were always positive echo generations regardless of the rephasing pulse area in the DRPE scheme if Gaussian light is involved. The resultant echo efficiency showed a damped oscillation with respect to the Gaussian rephasing peak-pulse area, where its maximum efficiency is 26% in amplitude. Considering potential eavesdropping in quantum channels due to device imperfectness-caused loopholes [29] and errors accumulated in each gate operation for quantum error-corrections [15] in fault-tolerant quantum computing [13,30], achieving near perfect retrieval efficiency is an essential requirement for quantum memory implementations. For a near maximal echo efficiency, the CDR echo protocol must be satisfied with spatially uniform rephasing pulses. The present discussion opens a door to practical applications of photon echoes for quantum information processing.

## ACKNOWLEDGMENTS

This work was supported by ICT R&D program of MSIP/IITP (1711028311: Reliable crypto-system standards and core technology development for secure quantum key distribution network).

Corresponding author's information: [*]bham@gist.ac.kr